# ESTIMATION AND UNCERTAINTY QUANTIFICATION OF YIELD VIA STRAIN RECOVERY SIMULATIONS

Paul N. Patrone

National Institute of Standards and Technology

100 Bureau Drive

Gaithersburg, MD 20899

## ABSTRACT

In computational materials science, predicting the yield strain of crosslinked polymers remains a challenging task. A common approach is to identify yield via the first critical point of stress-strain curves produced by molecular dynamics simulations. However, the simulated data can be excessively noisy, making it difficult to extract meaningful results. In this work, we discuss an alternate method for identifying yield on the basis of residual strain computations. Notably, the associated raw data produce a sharper signal for yield through a transition in their global behavior. As we show, this transition can be analyzed in terms of simple functions (e.g. hyperbolas) that admit straightforward uncertainty quantification techniques.

## 1. INTRODUCTION

In computational materials science, estimating the yield strain $\varepsilon_y$ of thermoset polymers remains a challenging task. Key difficulties arise from the general observations that: (I) these systems exhibit a continuous spectrum of relaxation times [1]; and (II) atomistic models are often necessary to capture the relevant physics of such relaxations [2]. As a result, it is becoming common for research groups to use molecular dynamics (MD) simulations in an effort to balance competing length and time-scale requirements [2].

While this approach offers a compromise that is often acceptable in R&D settings, it has nonetheless forced modelers to confront the inherent limitations of MD. In particular, high-throughput applications generally require the use of small (e.g. 5000 atom) systems, which exhibit large fluctuations in simulated data. In the case of yield, this means that standard estimation procedures (e.g. based on critical points of stress-strain curves) suffer from high levels of uncertainty that diminish the usefulness of the associated predictions (see, for example, Ref. [3]). This observation has led us to consider alternative methods of computing this quantity.

In this work, we propose to estimate yield strain through analysis of simulated residual-strain data. Analogous experimental results were obtained as far back as 1996 by Quinson *et al.*, who showed that residual strain of linear polymer chains (I) is zero up to yield, and (II) subsequently grows linearly with applied strain beyond yield [1]. Motivated by these results, we show how a global hyperbola analysis can be used to identify the onset of this linear behavior, and consequently yield. Moreover, we demonstrate how a bootstrap-style analysis of the resulting fit can be used to estimate uncertainties in the associated predictions, thereby quantifying our confidence in the simulations.

We emphasize that this analysis is limited to predicting yield and estimating uncertainties within the context of a single simulation. This is important insofar as finite-size and -time averaging can introduce an additional between-simulation uncertainty associated with under-sampling of crosslinked structures [4]. In the case of the glass-transition temperature, an analysis has been devised to quantify this additional "dark" uncertainty [4]. However, a comparable treatment for yield is complicated by the structure of the underlying stress and strain tensors. We leave further analysis for later work. Moreover, we do not rigorously pursue validation (or comparison with experiment), since open questions remain about verification (or estimation of uncertainties within the context of simulations alone).

## 2. OVERVIEW OF RESIDUAL STRAIN SIMULATIONS

In 1996, Quison et al. showed that deformation-relaxation experiments can be used to quantify the rate-dependence of relaxation modes in linear polymers such as polystyrene [1]. As a byproduct of this work, they generated plots of residual strain data $\varepsilon_r$ as a function of the applied strain ε, where

$$\varepsilon_r = \frac{l_f(\varepsilon) - l_0}{l_0}$$

$l_0$ is the initial length in the loading direction, and $l_f$ is the final length after applying a strain and then allowing the system to relax. Interestingly, they observed that yield (or the onset of plastic deformation) occurred at the first value of $\varepsilon$ for which the material exhibited a non-zero residual strain. Although not discussed by the authors, it is also noteworthy that in all of their results, $\varepsilon_r$ is approximately a linear function of the applied strain beyond yield. Critically, this observation holds irrespective of either the deformation rate, temperature, or relaxation time. Such results have since been experimentally reproduced for thermosets commonly found in aerospace applications [5].

Given the inherent length and time-scale limitations of MD, these observations are encouraging, since they suggest the possibility of a rate-independent method for determining yield *in silico*. We thus attempted to reproduce results in Ref. [5] using MD simulations of a roughly 5000 atom, 50/50 mixture of 4,4-diaminodiphenyl sulfone (44DDS) and digycidyl ether of Bisphenyl A (BisA), a two-functional epoxy. We refer to this system as 44BA. Details of the system preparation are provided in another manuscript [4,6], and we omit such a discussion here.

Figure 1 shows the results of a simulated residual-strain measurement, which is analogous to Fig. 4 in Ref. [5]. In order to generate this plot, we first strained the system by fixed, volume conserving increments at a variable rate determined by a convergence criterion on the running average stress; see Refs. [4,6] for details of how the convergence criterion works. After each strain increment, we saved the final structure for later analysis. Each simulation was a minimum of 20 ps long, with an average on the order of 60 ps to 80 ps. All strain simulations were performed using an NVT constraint with the Andersen thermostat at 300 K. Residual strains were then estimated (as a function of applied strain) by relaxing the saved unit cells with an NPT simulation and computing

$$\varepsilon_r = \sum_{i=1}^{3} \frac{|l_{f,i}(\varepsilon) - l_{0,i}|}{l_{0,i}}$$

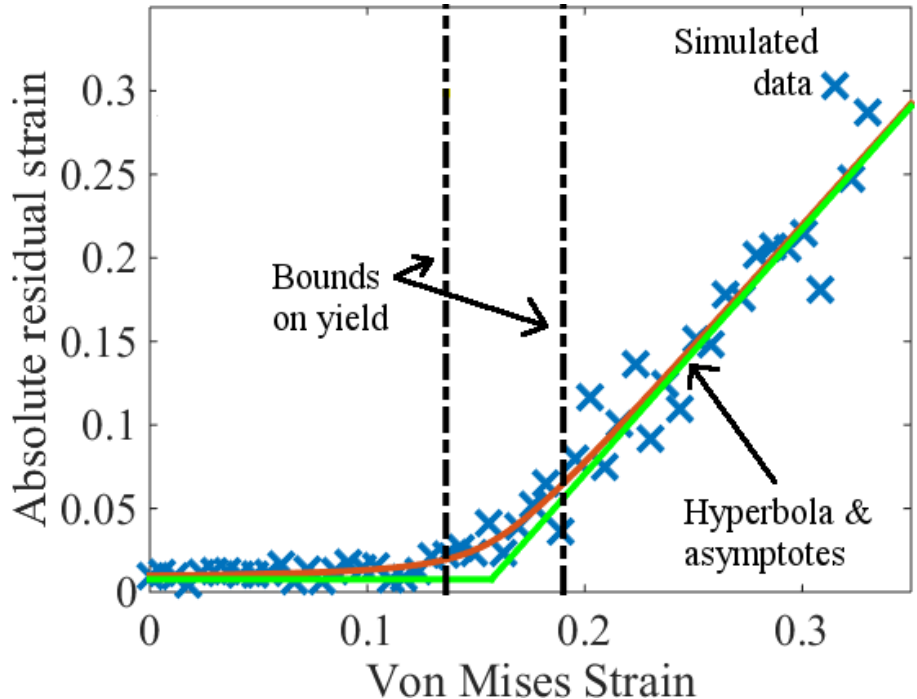

*Figure 1: Simulated residual strain data showing a hyperbola fit and bounds on yield. The bounds are taken as the minimum and maximum values of yield computed via the synthetic dataset approach described below.*

where $l_{f,i}$ and $l_{0,i}$ are the final and initial lengths of the $i^{th}$ side of the unit cell.[1] These latter simulations used the Parrinello barostat to ensure that the unit cells reached a well equilibrated, stress-free state. Analogous to before, a convergence criterion on the system dimensions was used to determine the simulation duration needed to reach an equilibrated state.

---

[1] We use a modified definition of residual strain (relative to Refs. [1] and [5]) because simulations provide additional information about deformation in three independent directions.

Several remarks are in order. Although noisy, the simulated residual strains show a bilinear character observed in experiments. In contrast to experiments, however, the simulated residual strains never fully go to zero for small applied strains. This occurs because the absolute values in our definition of $\varepsilon_r$ transform thermal fluctuations into one-sided noise in $\varepsilon_r$. Moreover, the transition in $\varepsilon_r$ associated with yielding is relatively smooth and occurs over a modest range of applied strains. Physically, we speculate that this arises from the fact that the anelastic relaxation mechanisms near yield have timescales that are poorly sampled by MD. In the next section, we show how hyperbola asymptotes can be used to estimate yield despite this lack of a sharp transition.

## 3. HYPERBOLA ANALYSIS AND UNCERTAINTY QUANTIFICATION

Given the data in Fig. 1, we estimate yield by first fitting a hyperbola $H$ to our simulated applied and residual strains, $\varepsilon_j$ and $\varepsilon_{r,j}$ (which are indexed by $j$). We find that it is convenient to use the parametrization

$$H = a + \frac{b}{2}(\varepsilon - \varepsilon_y) + \sqrt{\frac{b^2}{4}(\varepsilon - \varepsilon_y)^2 + e^c},$$

where $a$, $b$, $c$, and $\varepsilon_y$ are free parameters that are determined by a least-squares procedure. We generically denote this collection of parameters as $\varphi$. Given these, we identify the yield strain $\varepsilon_y$ with the hyperbola center, or equivalently the intersection of the hyperbola asymptotes. Physically we adopt the interpretation that these asymptotes characterize an "idealized" behavior of the simulation were it not to suffer from finite-size and -time effects.

In general, we find that a non-weighted least squares often gives reasonable estimates of yield, but not universally so. In particular, it is known that individual torsions in small-scale simulations can introduce large fluctuations into simulated quantities when a system is under high-strain [5]. Consequently, noise has a tendency to increase with the applied strain. In order to account for this, it is reasonable to determine $\varepsilon_y$ via a weighted least-squares fit of the data to a hyperbola. Figure 1 shows a fit obtained from the following iterative procedure: (I) compute an unweighted estimate of the hyperbola $H$; (II) estimate a power law $P(\varepsilon)$ for the residual data $H(\varphi, \varepsilon_j) - \varepsilon_{r,j}$; (III) compute a weighted least-squares estimate of $H$ with a weight-factor $1/P(\varepsilon)$. As the figure shows, this procedure allows for some flexibility in interpretation of the high-strain data while returning a reasonable estimate of yield.

To estimate uncertainties associated with our yield calculation, we perform a bootstrap-style analysis using repeated noise sampling of our model for the residual data. In particular, we define

$$\tilde{\varepsilon}_{r,j} = H(\varphi, \varepsilon_j) + \sqrt{P(\varepsilon_j)} N_j(0,1)$$

as a statistical model of residual strain data, where $P(\varepsilon_j)$ is determined according to the procedure described above, $j$ indexes strain increments, and $N_j(0,1)$ are uncorrelated Gaussian random variables with mean zero and variance 1. Realizations of $\tilde{\varepsilon}_{r,j}$ are inexpensive to compute using random number generators. Hence, we use this noise model to generate thousands of synthetic datasets, which in principle have the same statistical structure as the original dataset. Applying the hyperbola analysis to these datasets then generates a distribution of yield values associated with our uncertainty in the fit procedure; see Fig. 2.

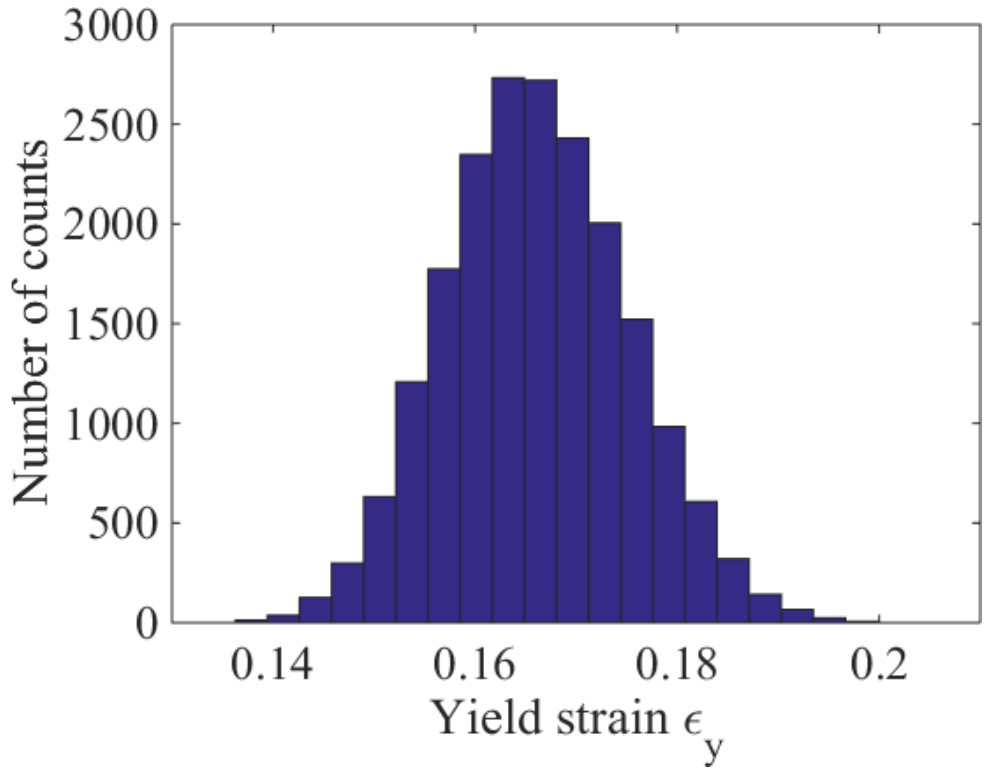

*Figure 2: Histogram of yield values computed from repeated noise sampling of synthetic datasets.*

## 4. CONCLUSIONS

Motivated by experimental work showing that the bilinear character of residual strain data is independent of strain rate, temperature, and relaxation times, we investigated how an analogous simulation protocol can be used to estimate yield. By analyzing the resulting data in terms of hyperbolas, we also showed how to (I) estimate the transition in residual strain that is associated with yield; and (II) quantify uncertainties in this procedure. We emphasize that in general, the methods discussed here only quantify uncertainties within the context of a single simulation. As previous work has shown, multiple simulations and comparison thereof may be necessary to construct a more complete picture of the computational predictions. However, qualitative agreement with experimental results warrants further investigation into the usefulness and validity of the method.

*Acknowledgements:* The author thanks Anthony Kearsley and Thomas Rosch for useful feedback during preparation of this manuscript. This work is a contribution of the National Institute of Standards and Technology and is not subject to copyright in the United States.